\documentclass[12pt,twoside,fleqn]{article}
\usepackage{epsfig}
\usepackage{psfig}
\def\lsim{\raise0.3ex\hbox{$<$\kern-0.75em\raise-1.1ex\hbox{$\sim$}}}
\def\gsim{\raise0.3ex\hbox{$>$\kern-0.75em\raise-1.1ex\hbox{$\sim$}}}
\makeatletter
\@addtoreset{equation}{section}
\makeatother

\newcommand{\beqn}{\begin{equation}}
\newcommand{\eqn}{\end{equation}}
\newcommand{\bqa}{\begin{eqnarray}}
\newcommand{\eqa}{\end{eqnarray}}
\newcommand{\bqas}{\begin{eqnarray*}}
\newcommand{\eqas}{\end{eqnarray*}}
\newcommand{\bdm}{\begin{displaymath}}
\newcommand{\edm}{\end{displaymath}}

\begin{document}
\thispagestyle{empty}
%
 \mbox{} \hfill CERN-TH/2000-204\\
 \mbox{} \hfill BI-TP 2000/22\\
 \mbox{} \hfill hep-ph/007093\\
\begin{center}
{{\large \bf Finite Temperature Meson Correlation Functions \\
in HTL Approximation} 
 } \\
\vspace*{1.0cm}
{\large F. Karsch$^{1,2}$, M. G. Mustafa$^{3,4}$ and M. H. 
Thoma$^{2,4}$\footnote{Heisenberg fellow}}

\vspace*{1.0cm}

{\normalsize
$\mbox{}$ {$^1$Fakult\"at f\"ur Physik, Universit\"at Bielefeld,
D-33615 Bielefeld, Germany}\\
$\mbox{}$ {$^2$Theory Division, CERN, CH-1211 Geneva 23, Switzerland} \\
$\mbox{}$ {$^3$Theoretical Nuclear Physics Division, Saha Institute of Nuclear 
Physics, 1/AF Bidhan Nagar, Calcutta - 700 064, India}\\  
$\mbox{}$ {$^4$Institut f\"ur Theoretische Physik, Universit\"at Giessen,
D-35392 Giessen, Germany}  

}
\end{center}
\vspace*{1.0cm}
\centerline{\large ABSTRACT}

\baselineskip 20pt

\noindent
We calculate temporal correlators and their spectral functions
with meson quantum numbers in the 
deconfined phase of QCD using the hard thermal loop (HTL) approximation
for the quark propagator.
Although this approach does not result in a complete next-to-leading order 
perturbative calculation it takes into account important
medium effects such as thermal quark masses and Landau 
damping in the quark-gluon plasma. 
We show that both effects lead to competing modifications of the free 
mesonic correlation
functions. We find that correlators in scalar channels are only
moderately influenced by the HTL medium effects, while the HTL-vertex 
corrections lead to divergent vector correlators. 
\vfill
\eject
\baselineskip 15pt

\section{Introduction}

While the lattice calculations of hadron properties in the vacuum have reached
quite satisfactory precision, little is known from such first principle
calculations about basic hadronic parameters in a thermal medium, e.g. masses and 
widths at finite temperature. Lattice calculations of such quantities at 
zero temperature generally proceed  through the calculation
of correlation functions in Euclidean time. This approach naturally carries 
over to the calculation of spatial correlation functions at finite temperature.
Such spatial correlation functions indeed show evidence for sudden changes of 
{\it in medium} hadron properties above $T_c$ \cite{DeT87}. 
However, they provide only 
indirect evidence for modifications of, e.g. hadron masses and their widths.
The appropriate approach here would be a detailed analysis of temporal
correlation functions \cite{Has93,Boy94}, which at finite temperature are 
restricted to the Euclidean time interval $[0,1/T]$. The interesting 
information on hadronic states is then encoded in the spectral functions for 
these correlators \cite{Shu93}. 

Currently available results from lattice calculations show significant 
changes in the behaviour of temporal correlation functions in the high 
temperature plasma phase of QCD \cite{Has93,Boy94,Boy95,deF99}. 
However, at least close to $T_c$ the correlation functions clearly deviate 
from those of freely propagating quarks. It thus is important to
understand in how far the temporal correlation functions carry
information about the existence or non-existence of bound states or 
resonances in the plasma phase. Various calculations within the framework 
of low energy effective models also suggest strong 
modifications of hadron properties \cite{models} and consequently also 
of the spectral functions \cite{Chi98}. However, it is difficult in such
model calculations to deal with the quark substructure of hadrons, which 
will become important at high temperature where one expects
to find indications for the propagation of almost free, massless quarks.
Eventually it is the hope, that spectral methods \cite{Jar96}, which
successfully have been applied to hadron spectral functions at
zero temperature \cite{Nak99}, can also be applied at finite temperature. 
In particular, in the high temperature limit, well above the QCD
phase transition temperature, it then might be appropriate to compare 
lattice calculations
for temporal hadron correlators also with perturbative calculations.
At least to some extend non-perturbative information can also be incorporated 
in such an analytic calculation by using the hard thermal loop (HTL) 
resummation scheme \cite{htl}. The recent successes in reproducing the QCD 
equation of state calculated on the lattice \cite{Boy96,Kar00} with 
HTL-resummed perturbative calculations for $T\; \gsim\; 2T_c$ 
\cite{Bla99,And99} 
suggest that this may be a reasonable starting point also for the 
description of other properties of the high temperature plasma phase
\cite{Tho98}. 
Indeed, the spectral functions, which one will extract from an analysis of 
temporal correlators, are closely related to quark-antiquark annihilation 
processes in the quark-gluon plasma. For the vector channel this is linked 
to the dilepton production at high temperature, which has been studied in the
HTL-approximation \cite{Bra90}. 
The temperature dependence of the pseudo-scalar correlator is related to the 
chiral condensate. Thermal effects on this as well as the pseudo-scalar
masses and dispersion relations influence the appearance or suppression
of a disoriented chiral condensate which might lead to observable effects 
in relativistic heavy ion collisions \cite{Sch99}.

We will analyze here the structure of temporal correlation functions
within the context of HTL-resummed perturbation
theory. In the infinite temperature limit the free field behaviour is
expected to give the dominant contribution to the spectral
functions for energies $\omega \sim T$ \cite{Ele88,Flo94}.  
Here we will use the HTL-resummed quark propagator \cite{Bra90,Kli82} 
for calculating the temporal correlators and their spectral functions.
In this way important in-medium properties of quarks in a quark-gluon plasma
are taken into account\footnote{It should be noted that we are not aiming 
at a calculation in the strictly perturbative sense, which holds only in
the weak coupling limit, $g\ll 1$. Rather we want to consider QGP medium 
effects using the HTL quark propagator for comparing with lattice 
results at realistic values of the coupling constant.}. 
As we will see, these medium
effects, will lead to competing effects in thermal 
correlation functions. On the one hand the effective quark propagator
takes into account the 
generation of thermal quark masses, $m_T \sim g(T) T$. This cuts off the 
low frequency part in the spectral functions and thus will lead to a steepening
of thermal correlation functions. On the other hand it also contains the
contributions from plasmino modes as well as interactions of quarks and 
antiquarks with gluons in the thermal heat bath (Landau damping). 
This enhances the
contribution of soft modes with $\omega \sim g(T) T$ which, in fact,  
will dominate the structure of spectral functions at low energies even at 
rather high temperature. 
These contributions from soft modes
will lead to a flattening of thermal correlation functions. We will discuss 
the interplay between both features of HTL-resummed correlation functions in
this paper. 

In the next section we will present the framework for the calculation
of thermal meson correlation functions in the HTL-approximation and give
results for the scalar and vector spectral function. In Section 3 we
compare the resulting thermal meson correlation functions with the 
leading order perturbative (free) correlators. Finally we give our 
conclusions in Section 4.

\section{Thermal Meson Correlation Functions}

\subsection{Definitions}

We want to analyze the behaviour of meson correlation functions in the high
temperature limit. They are constructed  from meson currents 
$J_M (\tau,\vec{x}) =\bar{q}(\tau, \vec{x})\Gamma_M q(\tau, \vec{x})$, 
where $\Gamma_M$ is an appropriate combination of $\gamma$-matrices
that fixes the quantum numbers of a meson channel; {\it i.e.,} $\Gamma_M =
1$, $\gamma_5$, $\gamma_\mu$, $\gamma_\mu \gamma_5$ for scalar,
pseudo-scalar, vector and pseudo-vector channels, respectively. 
The thermal two-point functions in coordinate space, $G_M(\tau,\vec{x})$,
are defined as 
\begin{eqnarray}
G_M(\tau,\vec{x}) &=& 
\langle J_M (\tau, \vec{x}) J_M^{\dagger} (0, \vec{0}) \rangle 
\nonumber \\
&=& T \sum_{n=-\infty}^{\infty} \int 
{{\rm d}^3p \over (2 \pi)^3} \;{\rm e}^{-i(\omega_n \tau- \vec{p} \vec{x})}\;
\chi_M(\omega_n,\vec{p})~~,
\label{tempcor}
\end{eqnarray}
where $\tau \in [0,1/T]$, and the Fourier transformed correlation function
$\chi_M(\omega_n,\vec{p})$ is given at the discrete Matsubara modes, 
$\omega_n = 2n \pi T$. The imaginary part of the momentum space correlator
gives the spectral function $\sigma_M(\omega,\vec{p})$,
\begin{equation}
\chi_M(\omega_n,\vec{p}) = -\int_{-\infty}^{\infty} {\rm d}
\omega\; {\sigma_M(\omega,\vec{p}) \over i\omega_n - \omega +i\epsilon}
\quad \Rightarrow \quad
\sigma_M(\omega,\vec{p}) = {1\over \pi} {\rm Im}\;  \chi_M(\omega,\vec{p})~~. 
\label{defspec}
\end{equation}

Using eqs.~\ref{tempcor} and \ref{defspec} we obtain
the spectral representation of the thermal correlation functions in
coordinate space at fixed momentum ($\beta =1/T$),    
\begin{equation}
G_M(\tau,\vec{p}) =  \int_{0}^{\infty} {\rm d} \omega\; 
\sigma_M (\omega,\vec{p})\; 
{{\rm cosh}(\omega (\tau - \beta/2)) \over {\rm sinh} (\omega \beta/2)}~~.
\label{speccora}
\end{equation}
For a perturbative analysis of these correlation functions at high 
temperature it is convenient to introduce dimensionless variables, 
$\tilde{\omega}=\omega/T$, $\vec{\tilde{p}}=\vec{p}/T$,
$\tilde{\tau}=\tau T$, the reduced spectral function 
$\tilde{\sigma} (\tilde{\omega},\vec{\tilde{p}}) \equiv 
\sigma (\omega,\vec{p})/T^2$, and the reduced correlator   
$\tilde{G}_M (\tilde{\tau},\vec{\tilde{p}}) \equiv G_M(\tau,\vec{p})/T^3$.

\subsection{Free Meson Spectral Functions}

The starting point for a calculation of the meson spectral functions
and the meson correlation functions is the momentum space representation
of the latter \cite{Flo94}. To leading order perturbation theory
one has to evaluate the self-energy diagram shown in Fig.~1a, 
\begin{figure}
\label{fig:bubble}
\begin{center}
~\epsfig{file=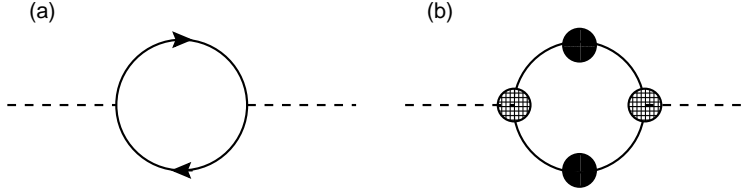,width=100mm}
\end{center}
\caption{The self-energy diagrams for free quarks (a) and in the
HTL approximation (b).}
\end{figure}
where the internal quark lines represent a bare quark propagator
$S_F(k_0, \vec{k})$ which can be expressed in terms of its spectral
function $\rho_F (\omega, \vec{k},m)$ and is conveniently written as
\begin{equation}
\hspace*{-1cm}
S_F(k_0, \vec{k}) = -(\gamma_0 \; k_0 - \vec{\gamma}\cdot \vec{k} + m )\;
\int_0^{1/T} \;{\rm d}\tau {\rm e}^{k_0\tau} \int_{-\infty}^{\infty}
{\;{\rm d}\omega}
\rho_{F} (\omega, \vec{k},m)\; [1-n_F(\omega)]\; {\rm e}^{-\omega \tau},  
\label{saclay}
\end{equation}
where $k_0=(2n+1)i\pi T$, $n_F(\omega) = 1/(1+\exp(\omega / T))$ and 
\begin{equation}
\rho_F (\omega, \vec{k}, m) = {1 \over 2\omega}\; 
(\delta (\omega - \omega_k) + \delta (\omega + \omega_k) )\quad ,
\label{free_rho}
\end{equation} 
with $\omega_k = \sqrt{\vec{k}^2 + m^2}$.
 
The thermal meson spectral functions are then obtained from eq.~\ref{defspec},
that is from the imaginary  part of the correlation functions in momentum space,
\begin{equation}
\chi_M(\omega, \vec{p}) = 2 N_c 
T \sum_n \int {{\rm d}^3k \over (2 \pi)^3} \;{\rm Tr}\biggl[ \Gamma_M
S_F(k_0,\vec{k}) \Gamma_M^{\dagger} 
S_F^{\dagger} (\omega - k_0, \vec{p}-\vec{k}) \biggr]
\label{momcor}
\end{equation}
In the case of free fermions this is easily evaluated. In the limit of 
vanishing external 
momentum one finds for the spectral functions,  
\begin{eqnarray}
\sigma_{\rm M}^{\rm free} (\omega, \vec{p}=0) &=& 
{N_c \over 4 \pi^2} \;
\Theta (\omega- 2m) \;  \omega^2  \;\tanh (\omega/4T)
\; \sqrt{1-\biggl({2m\over \omega}\biggr)^2 }
\nonumber \\
&~&\cdot \biggl(a_M + \biggl({2m \over \omega}\biggr)^2 b_M\biggr) 
\quad,\quad 
\label{spectral_sv}
\end{eqnarray} 
where different quantum number channels are characterized by the pair of parameters
$(a_M, b_M)$. For the scalar (s), pseudo-scalar (ps), vector (v) and pseudo-vector (pv)
channels they are given by (-1,1), (1,0), (2,1) and (-2,3), 
respectively\footnote{In the vector and pseudo-vector cases
we denote by $\sigma_M$ the trace over the Lorentz indices of
$\sigma_M^{\mu \nu}$.}.  
In the massless limit the spectral functions are chirally symmetric,
$|\sigma_{ps}| = |\sigma_{s}|$ and $|\sigma_{pv}| = |\sigma_{p}|$.
In this case the remaining integral in eq.~\ref{speccora} can  
be done analytically and one obtains for example in the pseudo-scalar case
\cite{Flo94},
\begin{equation}
\tilde{G}_{ps} (\tilde{\tau},\vec{p}=0) = 2\pi N_c\; ( 1-2\tilde{\tau})
{1+\cos^2(2\pi\tilde{\tau} ) \over \sin^3 (2\pi\tilde{\tau} )} 
+ 4 N_c \; {\cos(2\pi\tilde{\tau} ) \over \sin^2 (2\pi\tilde{\tau} )}\quad .
\label{cor_free_ps}
\end{equation}

\subsection{HTL-Approximation for Meson Spectral Functions}


Now we want to go beyond the free quark approximation and consider 
in-medium quark
propagators. A consistent way in the weak coupling limit ($g\ll 1$)
is the use of a 
HTL-resummed quark propagator (and quark-meson vertex) if the quark momentum
is soft, i.e. of order $gT$. These corrections to the spectral function are
of the same order as the free spectral function for small energy $\omega$.

Using the HTL resummation technique
important medium effects of the quark-gluon plasma such as effective
quark masses and Landau damping are taking into account. 
The HTL-resummed fermion propagator is obtained from eq.~\ref{saclay}
by replacing the free spectral function $\rho_F$ with the HTL-resummed
spectral function which for massless quarks is given by \cite{Bra90,Kli82} 
\begin{equation}
\rho_{\rm HTL} (k_0, \vec{k}) =  
{1\over 2} \rho_{+} (k_0, k)(\gamma_0 - i\; \hat{k}\cdot \vec{\gamma} ) 
+{1\over 2} \rho_{-} (k_0,k)(\gamma_0 + i\; \hat{k}\cdot \vec{\gamma} ) 
\label{htl_rho}
\end{equation}
with $\hat{k} = \vec{k}/k$, $k=|\vec{k}|$, and
\begin{eqnarray} 
\hspace*{-0.5cm}
\rho_{\pm} (k_0, k) \hspace*{-0.3cm}&=& \hspace*{-0.3cm}
{k_0^2 - k^2 \over 2 m_T^2} 
[\delta (k_0 - \omega_{\pm}) + \delta (k_0 + \omega_{\mp})]
+\beta_{\pm} (k_0, k) \Theta(k^2 -k_0^2) \label{htl_spec}\\
\hspace*{-0.5cm}\beta_{\pm} (k_0, k) \hspace*{-0.3cm}&=& \hspace*{-0.3cm}
-{m_T^2 \over 2} { \pm k_0 - k \over
\biggl[k(-k_0 \pm k) + m_T^2 \biggl( \pm 1 - {\pm k_0 -k \over 2k}
\ln{k+k_0 \over k-k_0} \biggr)\biggr]^2 + \biggl[ {\pi \over 2}
m_T^2 {\pm k_0 -k \over k} \biggr]^2 }\nonumber
\end{eqnarray}
Here $\omega_{\pm} (k)$ denote the two dispersion relations of quarks
in a thermal medium \cite{Bra90,Kli82} and $m_T=g(T)T/\sqrt{6}$ is the
thermal quark mass. We note that in addition to the appearance of two
branches in the thermal quark dispersion relation
the HTL-resummed
fermion propagator also receives a cut-contribution below the light-cone
($k_0^2<k^2$), which results from
interactions of the valence quarks with gluons in the thermal medium
(Landau damping). 
Furthermore, an explicit temperature dependence only
enters through $m_T(T)$. Also the HTL-resummed quark spectral function
can thus be written in terms of dimensionless, rescaled variables, e.g. 
$\tilde{\omega}=\omega /T$ etc. and the reduced meson spectral functions
$\tilde{\sigma}^{\rm HTL} = \sigma^{\rm HTL}/T^2$ will depend on temperature
only through $\tilde{m}_T = g(T)/\sqrt{6}$.
It also should be noted that the HTL resummed quark propagator
is chiral symmetric in spite of the appearance of an effective quark mass
\cite{Kli82}. In the following we thus will ignore the parity of the
meson states and will generically talk about scalar and vector channels
only\footnote{We will, however, show results for pseudo-scalar and
vector spectral functions and correlators which in our notation are
strictly positive.}.  

Inserting eq.~\ref{saclay} for $m=0$ together with eq.~\ref{htl_rho} 
and eq.~\ref{htl_spec} into eq.~\ref{momcor} we can determine the 
spectral functions for mesons in the HTL-approximation. In the vector
channel this also requires additional modifications of the vertex 
functions $\Gamma_M$, {\it i.e.,} the use of a HTL quark-meson vertex, 
as discussed in detail in \cite{Bra90}.
In the case of the scalar and pseudo-scalar spectral function the 
vertices are given by the bare vertices $\Gamma_s = 1$ and 
$\Gamma_{ps}=\gamma_5$, since contributions from a HTL resummation to
the vertices are suppressed in this case, {\it i.e.} lead to higher order 
corrections, as discussed for the scalar case
in the Yukawa theory \cite{Tho95}
and scalar QED \cite{Kra95}. In the case of QCD the absence of HTL 
corrections for scalar as well as pseudo-scalar vertices has been shown 
in the context of kinetic theory \cite{Bla94}.
If, on the other hand,
the bare vertex is proportional to $\gamma_\mu$ as in the vector meson 
case, the HTL vertex cannot be neglected, since in a gauge theory it is 
related to the HTL fermion propagator by Ward identities \cite{Bra90a}. 

The pseudo-scalar spectral function can then be written as
\begin{eqnarray}
\hspace{-1.0cm}\sigma_{\rm ps} (\omega,\vec{p})
\hspace{-0.3cm} &=& \hspace{-0.3cm}
2N_c ({\rm e}^{\omega/T} -1) \int {{\rm d}^3k \over (2\pi )^3} \; 
\int_{-\infty}^{\infty} {\rm d}x {\rm d}x'
\; n_F(x) n_F(x') \delta(\omega -x-x')\nonumber \\
&~&\cdot \biggl\{ (1-\vec{q}\cdot \vec{k})
[\rho_+(x,k) \rho_+(x',q) + \rho_-(x,k) \rho_-(x,q)]\nonumber \\
&~&+(1+\vec{q}\cdot \vec{k})
[\rho_+(x,k) \rho_-(x',q) + \rho_-(x,k) \rho_+(x,q)]\biggr\},
\label{sig_gen}
\end{eqnarray}
where $\vec{q}=\vec{p}-\vec{k}$. The corresponding relation for the
vector spectral function, which also includes HTL-vertex contributions,
is related to the dilepton production rate calculated
in Ref.~\cite{Bra90} in the HTL approximation, 
\begin{equation}
{\sigma_{\rm v}} (\omega , \vec{p}=0) = \frac{18\pi^2N_c}{5\alpha^2}
\> \left ({\rm e}^{\omega /T}-1\right )\> \omega^2\> \frac{{\rm d}W}
{{\rm d} \omega {\rm d}^3p}(\vec{p}=0)~. 
\label{dilep}
\end{equation}
Here $\alpha $ is the electromagnetic fine structure constant.

As the thermal meson correlation functions are constructed from two quark
propagators, they will receive pole-pole, pole-cut and cut-cut contributions,
{\it i.e.,}
the mesonic spectral functions for $\vec{p}=0$ are generically given by
\begin{equation}
\sigma^{\rm HTL}  (\omega) = \sigma^{\rm pp} (\omega) +  \sigma^{\rm pc} (\omega) +  
\sigma^{\rm cc} (\omega)  \quad.
\label{htl_sigma}
\end{equation}
Explicit expressions for the three different contributions to the 
pseudo-scalar 
spectral function, $\sigma^{\rm HTL}_{\rm ps}$, are given in the Appendix. 
Similar results for the vector spectral function $\sigma^{\rm HTL}_{\rm v}$ 
have been derived in \cite{Bra90} where also a detailed discussion of the 
physical processes related to the pole-pole, pole-cut and cut-cut contributions
is given. 
In particular, there are characteristic peaks that show up in the pole-pole 
contribution (van Hove singularities). They are caused by a diverging density 
of states which is inversely proportional to the derivative of the dispersion 
relations, $\omega'_{\pm}(k)$, appearing in eq.~A.1. Owing to the minimum in 
the plasmino branch\footnote{In Ref.\cite{Pes00} it has been 
argued that the full in-medium quark propagator leads in general to two 
branches in the dispersion relation, of which one exhibits a minimum.}  
these derivatives vanish at $\omega = 0.47 m_T$ and $1.856 m_T$.
Apart from values close to the van Hove singularities in $\sigma^{\rm pp}$ 
one finds that the cut contributions dominate the spectral function for small 
values of $\tilde{\omega}$, e.g.  for $\tilde{\omega}\; \lsim \; g(T)$.

In Fig.~2 we show the pole $(\tilde{\sigma}^{\rm pp})$ and 
cut $(\tilde{\sigma}^{\rm pc},~\tilde{\sigma}^{\rm cc})$ contributions to 
the scalar (Fig.~2a) and vector (Fig.~2b) spectral functions for 
the case $\tilde{m}_T=1$, extrapolating the HTL results, obtained in the 
weak coupling limit, to $g=\sqrt{6}$. Of course it is a priori not clear 
whether one can apply the HTL resummation method 
to such large couplings, although the extrapolation to realistic values  
of the coupling constant has been used for
various observables of the QGP \cite{Tho95a}. However, in
this way we can study the influence of medium effects at least
qualitatively and as will become clear, the small corrections found by 
us do justify our choice for the couplings.
\begin{figure}
\label{fig:sigma}
\begin{center}
\epsfig{file=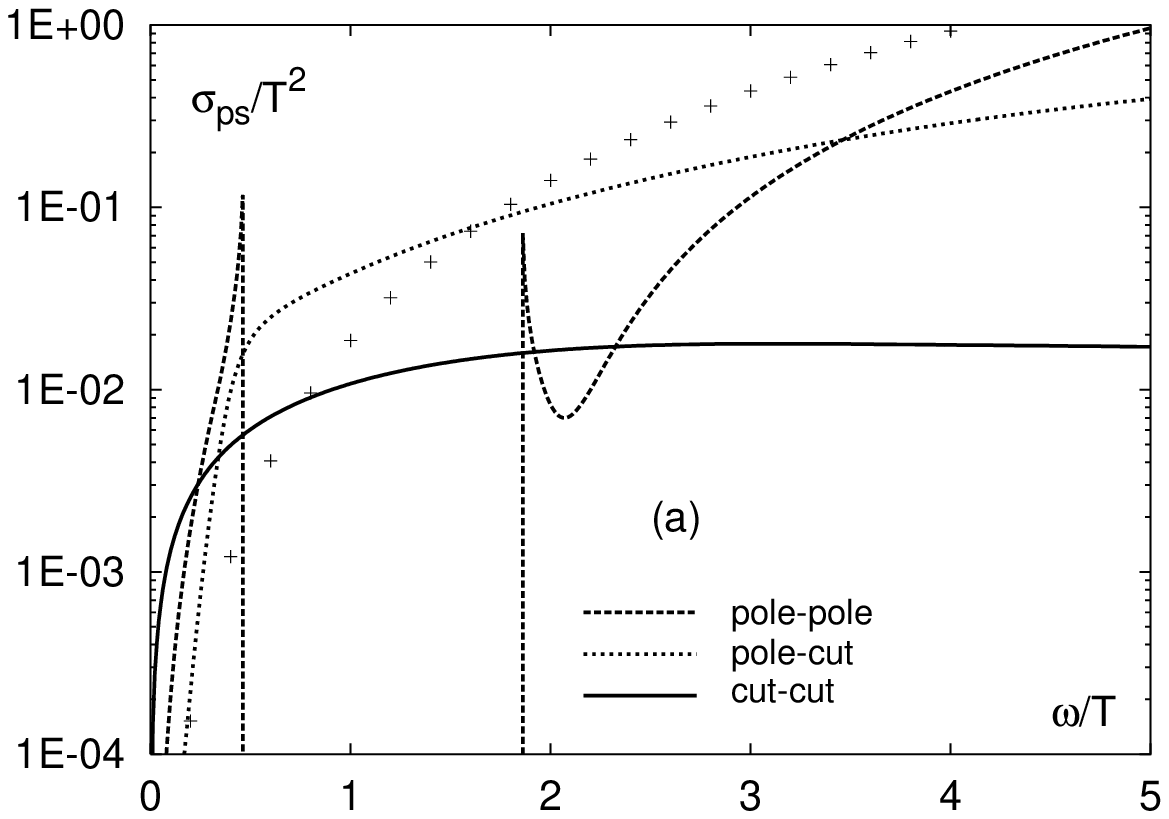,width=73mm}
\epsfig{file=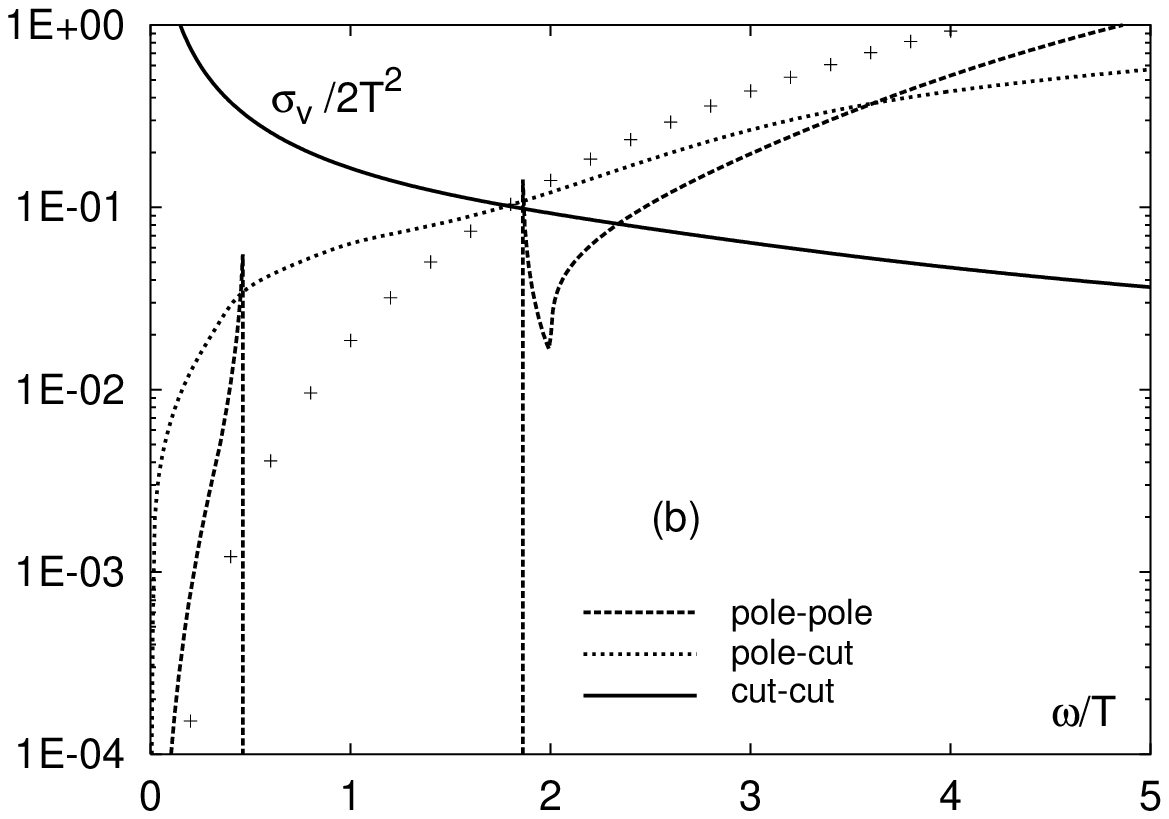,width=73mm}
\end{center}
\caption{The pole-pole, pole-cut and cut-cut contributions to the 
pseudo-scalar (a) and vector (b) spectral function for $\tilde{m}_T=1$.
The crosses show the free meson spectral function.
} 
\end{figure}
As can be seen in Fig.~2 the pole and cut contributions 
influence the spectral function in different ways. The former gives the 
dominant contribution for
large $\tilde{\omega}$. The deviations of $\sigma^{\rm HTL}$ from the free 
spectral function in this energy regime as well as the threshold for
$\tilde{\omega} \simeq 2~\tilde{m}_T$ is due to the presence of a 
non-vanishing thermal mass in the quark dispersion
relation and reflects the almost free
propagation of two quarks in the plasma. Additional interactions of these
quarks with the thermal medium (Landau damping)
are represented by the cut contributions. These lead to an enhancement 
over the free spectral functions for small values of $\tilde{\omega}$ as 
discussed above. Furthermore, we note that the pole-pole and pole-cut 
contributions to the spectral functions are similar in the scalar and vector 
channels. The cut-cut contribution, however, behaves differently at small
energies. While it vanishes for small $\tilde{\omega}$ in the scalar channel 
it diverges linearly in the vector channel.
This can be traced back to the structure of the effective
HTL-vertex, which contains a collinear singularity \cite{Bai94}. As a 
consequence of this singularity infinitely 
many higher order diagrams in the HTL 
expansion contribute to the same order in the coupling constant \cite{Aur98}.
This, of course, indicates that the low frequency part of the vector spectral
functions is inherently non-perturbative.

\section{HTL-Approximation for Thermal Meson Correlators}

\subsection{Thermal Pseudo-Scalar Meson Correlation Function}

Now we will compute the temporal correlators from the spectral functions,
derived in the last section, using (\ref{speccora}). Although the temporal 
correlators are dominated by hard energies $\omega \sim T$, we 
use the spectral functions, calculated within the HTL resummation scheme,
in order to take into account medium effects from the quark propagator
at least semi-empirically. 

The competing influence of pole and cut contributions to the HTL-resummed 
spectral functions carries over to the behaviour of thermal meson correlation 
functions. The appearance of a non-vanishing thermal quark mass tends to lead 
to a more rapid decrease of the correlator in Euclidean time than this is
the case for the free massless correlator. The enhancement of the low energy 
part which is due to the cut contributions, on the other hand, will counteract 
this trend. 
This is evident from the behaviour of thermal meson correlation functions 
in the scalar channel which is shown in Fig.~3 for $\tilde{m}_T=1$ and 2.   
\begin{figure}
\label{fig:meson}
\begin{center}
\epsfig{file=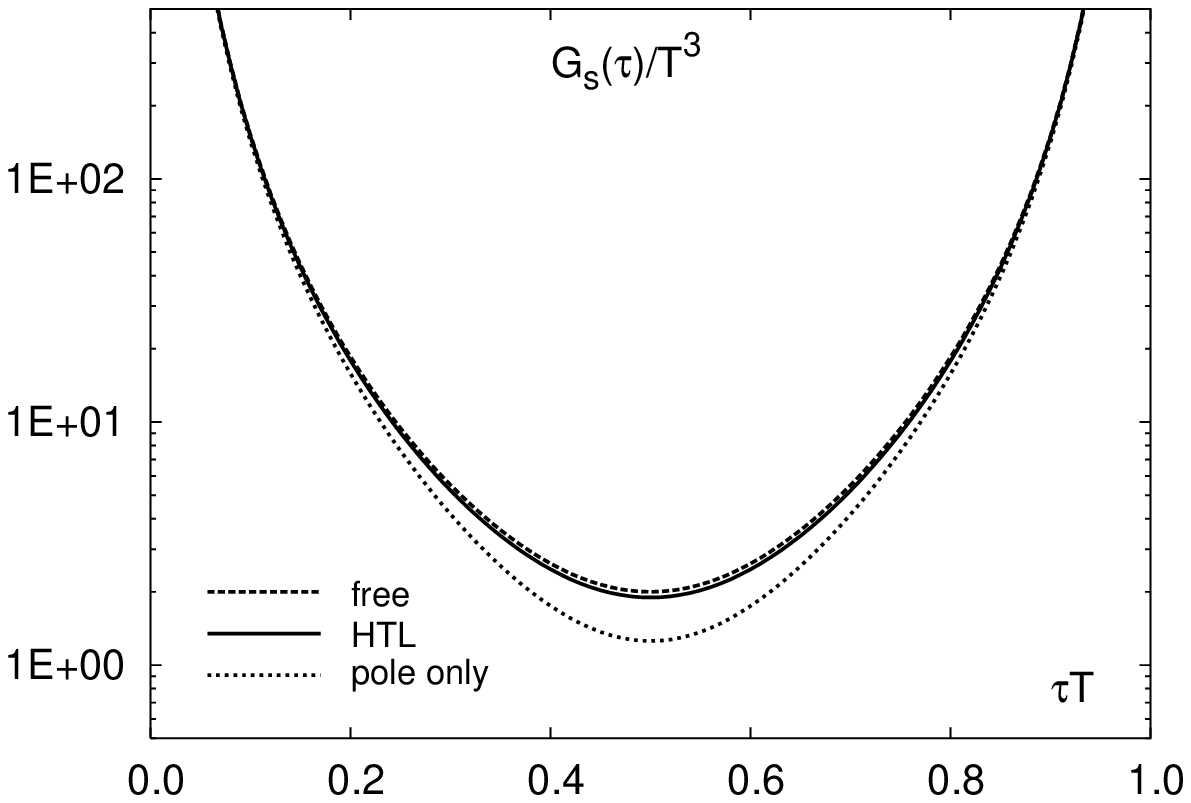,width=74mm}
\epsfig{file=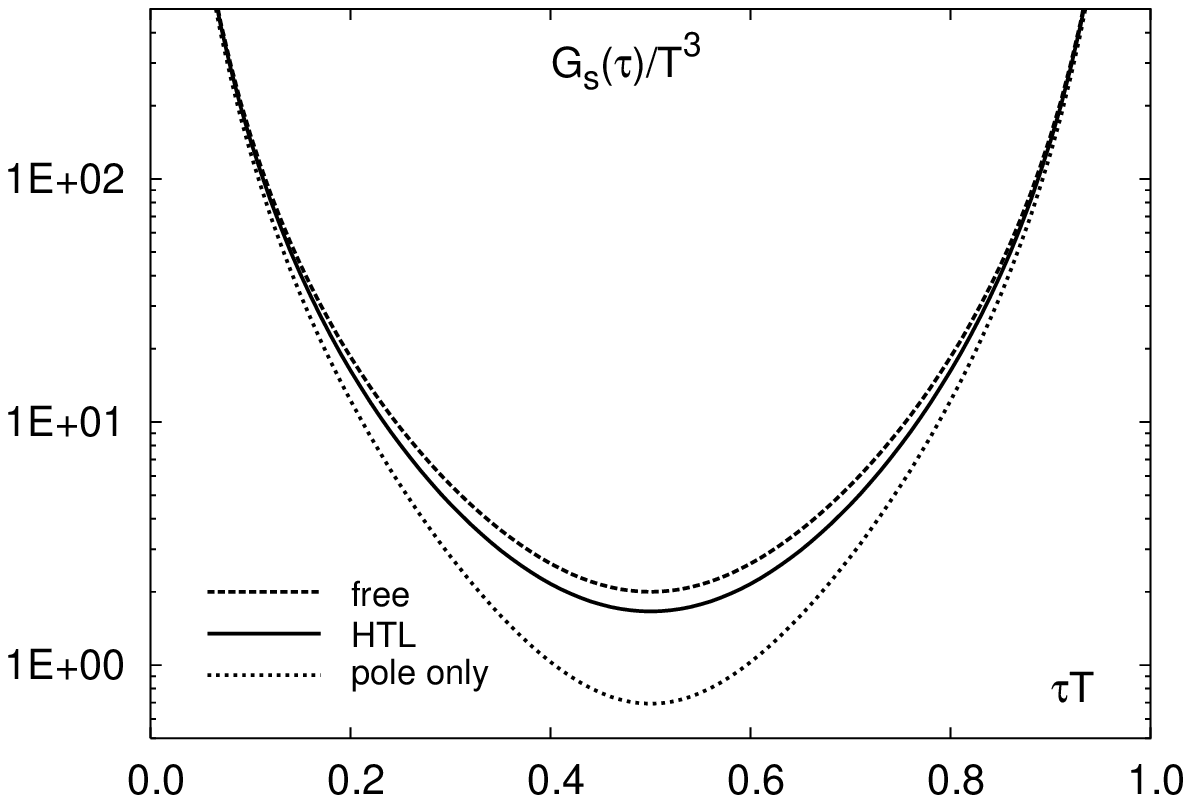,width=74mm}
\end{center}
\caption{The thermal pseudo-scalar meson correlation function in the HTL 
approximation for $\tilde{m}_T=1$ (left) and $\tilde{m}_T=2$ (right).  
The curves shown the 
complete thermal correlator (middle line), the correlator constructed from 
$\sigma^{\rm pp}_{\rm s}$ only (lower line) and the free thermal correlator 
(upper line).}
\end{figure}
As expected the correlator constructed from the pole-pole 
contribution alone is 
steeper than the free correlator and, moreover, is strongly dependent on
$\tilde{m}_T$. The cut contributions, however, enhance the 
low energy contributions in the spectral function and thus flattens the 
correlator again. Somewhat surprisingly for $\tilde{m}_T \simeq 1$ this seems 
to compensate almost completely the deviations from the free correlator 
introduced by the pole contributions. The difference between the free and
HTL-resummed correlators is largest for $\tau T \simeq 1/2$ where the 
contribution from the low energy regime in the spectral function is largest. 
For $\tau T \rightarrow 0$, and $\tau T \rightarrow 1$, on the other hand, the 
free and HTL-resummed correlators approach each other as 
$\lim_{\omega\rightarrow \infty} \sigma^{\rm HTL}(\omega)/\sigma^{\rm free} 
(\omega) = 1$.  These features are amplified in the ratio 
$G_{ps}^{\rm HTL} (\tau) / G_{ps}^{\rm free} (\tau)$ which is shown in Fig.~4.
\begin{figure}
\label{fig:ratio}
\begin{center}
\epsfig{file=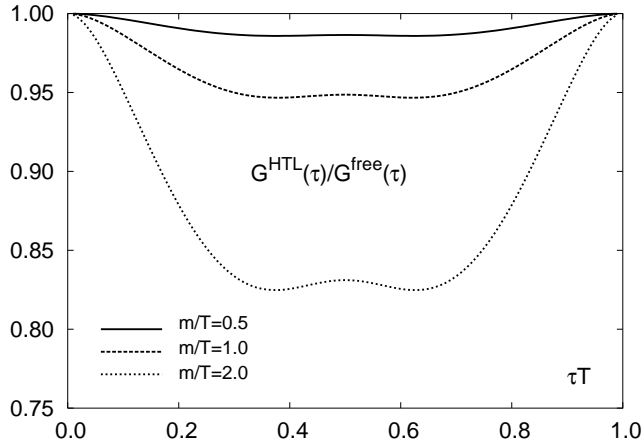,width=90mm}
\end{center}
\caption{The ratio of the HTL-resummed and free thermal pseudo-scalar 
correlation function versus Euclidean time $\tau$ in units of the 
temperature. Shown are results for thermal quark masses 
$\tilde{m}_T=0.5$ (top), $1.0$ (middle) and  $2.0$ (bottom).}
\end{figure}

\subsection{Thermal Vector Meson Correlation Function}

As already discussed in section 2.1 the calculation of the vector 
correlators within the HTL method requires the use of effective quark-meson 
vertices as shown in Fig.~1b. This does lead to
a linear divergence of the spectral function in the vector channel at 
low frequencies, which
in turn renders the temporal correlator infrared divergent. In fact, 
although the scalar correlation functions are infrared
finite, it is to be expected, that also in this case the low frequency
part of the HTL-resummed spectral functions will be modified 
significantly from contributions of higher order diagrams. It thus 
seems to be reasonable to consider modified correlation functions,
which are less sensitive to details of the low frequency part of the
spectral functions. We therefore define the subtracted correlators

\begin{equation}
\Delta \tilde{G}_M (\tau) \equiv \tilde{G}_M (\tau) - \tilde{G}_M (\beta/2)~.
\label{subG}
\end{equation}
In the subtracted correlation functions the infrared divergences are
eliminated. They are well-defined in the scalar as well as in the vector 
channels. In Fig.~5 we compare the HTL-resummed subtracted correlation 
functions with corresponding results for the free case. This shows that
after elimination of the infrared divergent parts the structure
of the pole and cut contributions is similar in scalar and vector
channels. The vector correlator seems to be even closer to the
leading order perturbative (free) correlator than the scalar correlation
function. 
 
\begin{figure}
\label{fig:subtracted}
\begin{center}
\epsfig{file=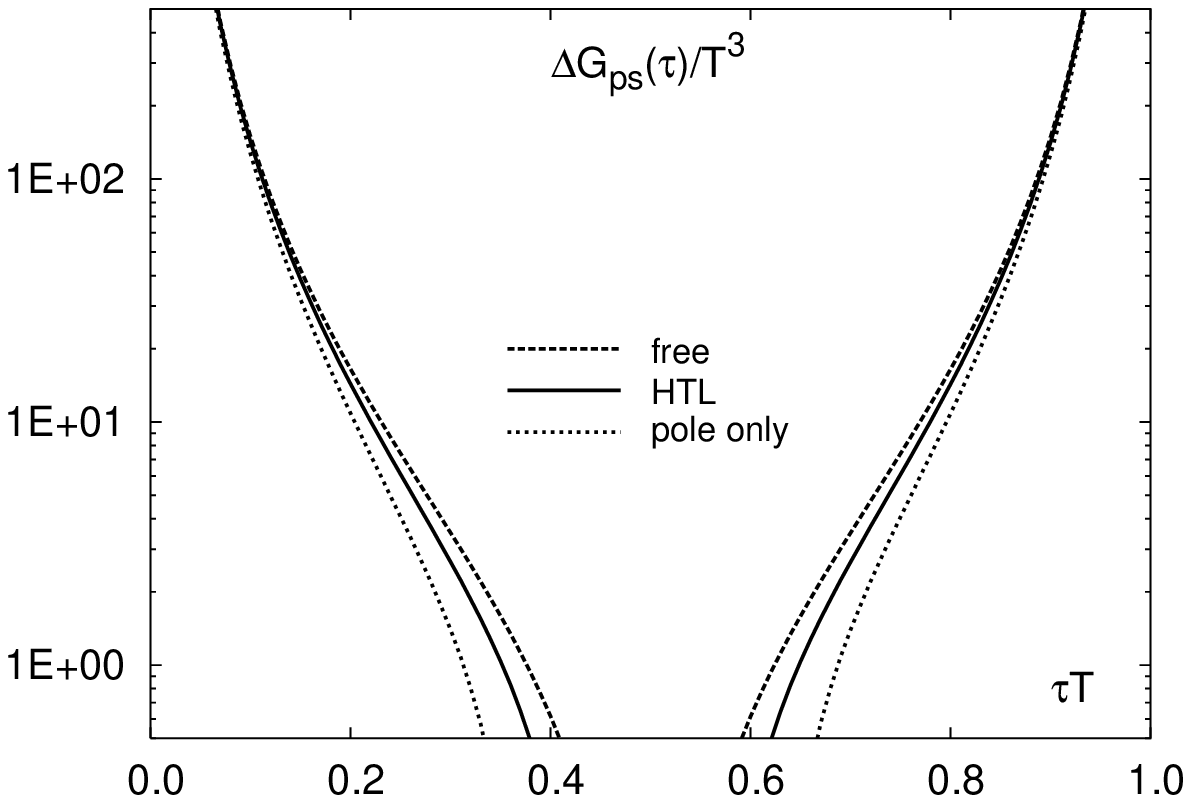,width=74mm}
\epsfig{file=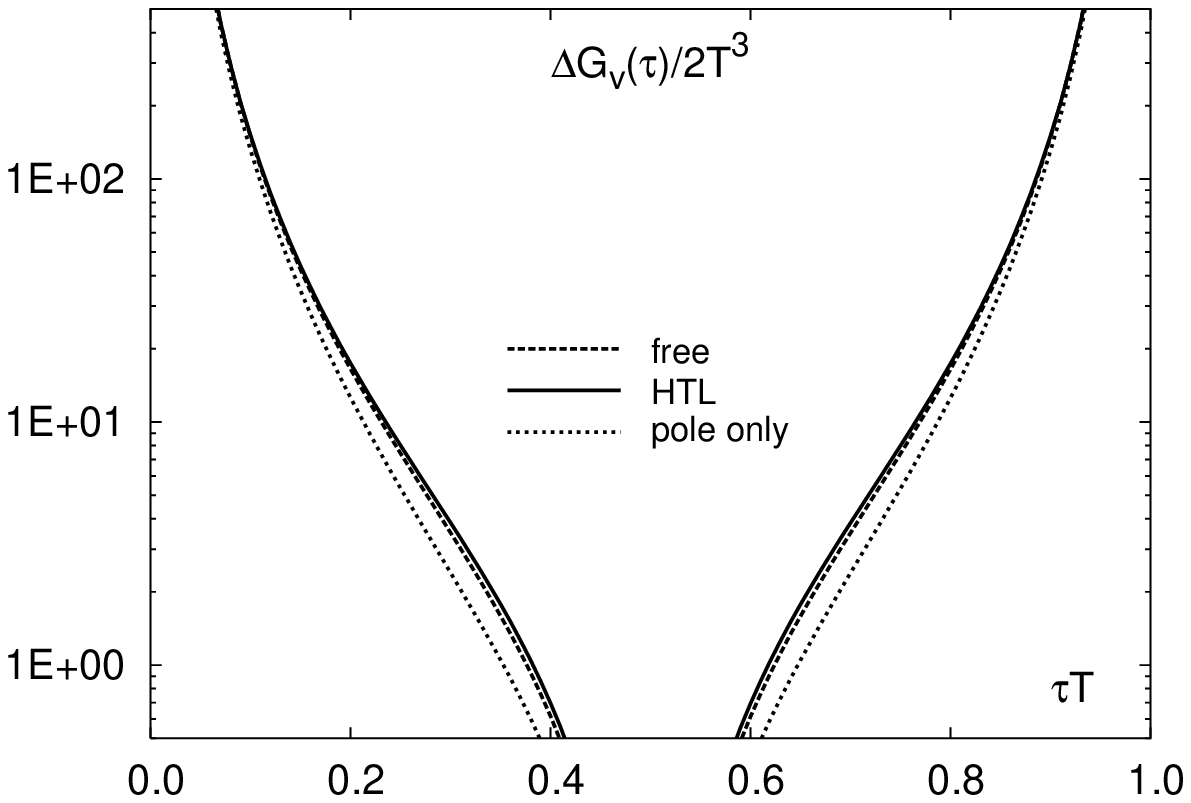,width=74mm}
\end{center}
\caption{The subtracted thermal pseudo-scalar (left) and vector (right) meson 
correlation functions in the HTL 
approximation for $\tilde{m}_T=2$.  The curves shown the 
complete thermal correlator (middle line), the correlator constructed from 
$\sigma^{\rm pp}_{\rm ps}$ only (lower line) and the free thermal correlator 
(upper line).}
\end{figure}

\section{Conclusions}

We have calculated thermal meson correlation functions 
and their spectral functions in the 
HTL-approximation; i.e., we have used the HTL-resummed quark propagator
(and HTL quark-meson vertex in the vector case) for considering important
in-medium properties of quarks in a QGP. 

We have analyzed the influence of the
various contributions to the HTL-resummed scalar and vector meson
spectral functions on the structure of the thermal correlators.
We generally find that the correlators in HTL-approximation --
after subtracting the infrared singularity in the vector correlator
-- are quite similar to those calculated in leading order perturbation theory
which correspond to free correlation functions.
Non-perturbative features of the HTL-resummed quark propagators such as
the generation of a thermal quark mass and Landau damping are clearly
visible in the meson spectral functions. However, they lead to competing
effects in the correlation functions and to a large extent compensate
each other. 

The main difference between the scalar and the vector channel using 
the HTL approximation is the different behaviour of the cut-cut contribution 
to the spectral functions at small energies. The vector spectral function 
diverges in the infrared limit leading to a singular expression for the
vector correlation function. This feature is in agreement with the 
observation, that
the dilepton production rate, which is closely related to the vector 
spectral function, cannot be computed within the HTL improved perturbation 
scheme for small invariant masses $M\simeq g^2T$.

The existing lattice calculations of thermal meson correlation functions
show that the correlators deviate from the free field result significantly
for temperatures $T\lsim 2T_c$. However, also at larger temperatures
it seems that the  scalar correlator only slowly approaches the free
correlation function and still differs in shape from the vector correlator.
This suggests that HTL-resummed perturbation theory, which gave a 
satisfactory description of bulk thermodynamics above $2T_c$, will not be 
appropriate for a quantitative analysis of thermal meson correlation 
functions at least in the pseudo-scalar case. In other words, 
the HTL medium effects (thermal quark masses, Landau damping) 
are not sufficient to explain the deviations of the pseudo-scalar correlator
from the free one as observed in lattice calculations. Therefore additional
non-perturbative effects, maybe related to chiral symmetry restoration,
appear to be important in the pseudo-scalar channel. 

\vspace{0.5cm}
\noindent
{\bf Acknowledgements:} 

\medskip
\noindent
The work of FK has been supported by the TMR network ERBFMRX-CT-970122 and the 
DFG under grant Ka 1198/4-1. MGM would like to acknowledge support from
AvH Foundation as part of this work was initiated during his stay at the 
University of Giessen as an Humboldt Fellow.

\vspace{0.5cm}
\noindent
{\Large \bf Appendix} 

\medskip
\noindent
We will give here explicit expressions for the pole-pole, pole-cut and cut-cut 
contributions to the pseudo-scalar spectral function. 
The pole-pole contribution ($\vec{p}=0$) is given by
\begin{eqnarray}
\hspace{-1.0cm}\sigma^{\rm pp}_{\rm ps} (\omega)\hspace{-0.3cm} &=&\hspace{-0.3cm} 
{N_c\over 2\pi^2} m_T^{-4} ({\rm e}^{\omega/T} -1)
\biggl[ n_F^2(\omega_+(k_1)) \bigl( \omega_+^2(k_1) -k_1^2 \bigr)^2 {k_1^2\over 2| \omega_+'(k_1)|}
\nonumber \\
&+&\hspace{-0.3cm}2 \sum_{i=1}^2
n_F(\omega_+(k_2^i)) \bigl[1-n_F(\omega_-(k_2^i))\bigr] \bigl(\omega_-^2(k_2^i) -(k_2^i)^2 
\bigr) \bigl(\omega_+^2(k_2^i) -(k_2^i)^2 \bigr) \cdot\nonumber \\
&~&\hspace{1.5cm} \cdot {(k_2^i)^2\over | \omega_+'(k_2^i)-\omega_-'(k_2^i)|}
\nonumber \\
&+&\hspace{-0.3cm} \sum_{i=1}^2
n_F^2(\omega_-(k_3^i)) \bigl(\omega_-^2(k_3^i) -(k_3^i)^2 \bigr)^2 
{(k_3^i)^2\over 2| \omega_-'(k_3^i)|}\; \biggr] \quad.  \qquad\qquad
\qquad\qquad\hfill  (A.1) \nonumber
\end{eqnarray}
Here $\omega_\pm (k)$ denote the quark dispersion relations for the ordinary 
quark (+) and the plasmino (-) branch \cite{Bra90}, 
$k_1$ is the solution of $\omega - 2 \omega_+(k_1) = 0$, $k_2^i$ and $k_3^i$
are the solutions of  $\omega - \omega_+(k_2^i)+\omega_-(k_2^i) =0$ and  
$\omega - 2  \omega_-(k_3^i) = 0$, respectively. Note that for small momenta the last two
equations can each have two solutions. Furthermore,  
$\omega_\pm'(k)\equiv ({\rm d}  \omega_\pm (x) / {\rm d} x)|_{x=k}$.
For the pole-cut contribution we find
\begin{eqnarray}\hspace{-1.0cm}\sigma^{\rm pc}_{\rm ps} (\omega)\hspace{-0.3cm} &=&
\hspace{-0.3cm}
{2N_c\over \pi^2} m_T^{-2} ({\rm e}^{\omega/T} -1) \int_0^\infty {\rm d}k\; k^2 \cdot \nonumber\\
&\cdot&\hspace{-0.3cm}\biggl[
~~~\Theta (k^2-(\omega-\omega_+)^2) n_F(\omega-\omega_+) n_F(\omega_+) \beta_+(\omega-\omega_+,k)
(\omega_+^2-k^2) \nonumber\\
&~&\hspace{-0.3cm}~+
\Theta (k^2-(\omega-\omega_-)^2) n_F(\omega-\omega_-) n_F(\omega_-) \beta_-(\omega-\omega_-,k)
(\omega_-^2-k^2) \nonumber\\
&~&\hspace{-0.3cm}~+
\Theta (k^2-(\omega+\omega_-)^2) n_F(\omega+\omega_-) [1-n_F(\omega_-)] \beta_+(\omega+\omega_-,k)
(\omega_-^2-k^2) \nonumber\\
&~&\hspace{-0.3cm}~+
\Theta (k^2-(\omega+\omega_+)^2) n_F(\omega+\omega_+) [1-n_F(\omega_+)] \beta_+(\omega+\omega_+,k)
(\omega_+^2-k^2) \biggr]\nonumber\\
& &\nonumber  \hspace{10.0cm}\qquad\hfill  (A.2)
\end{eqnarray}
Finally we obtain for the cut-cut contribution
\begin{eqnarray}\hspace{-1.0cm}\sigma^{\rm cc}_{\rm ps} (\omega)\hspace{-0.3cm} &=&
\hspace{-0.3cm}
{2N_c\over \pi^2} ({\rm e}^{\omega/T} -1) \int_0^\infty {\rm d}k\; k^2 \int_{-k}^{k}
{\rm d}x \; n_F(x) n_F(x-\omega) \Theta(k^2-(x-\omega)^2) \cdot \nonumber \\
&~&\hspace{3.0cm}\cdot \biggl[
\beta_+(x,k) \beta_+(\omega-x,k) + \beta_-(x,k) \beta_-(\omega-x,k) \biggr]\nonumber \\
& &\qquad \nonumber \hspace{9.0cm}\qquad\hfill (A.3)
\end{eqnarray}
The resulting contributions to the thermal correlator are then given by eq.~\ref{speccora}.

\end{document}